\documentclass[prd,nofootinbib,superscriptaddress,preprint]{revtex4}
\usepackage[T1]{fontenc}
\usepackage{amsmath,amssymb}
\usepackage{epsfig}
\usepackage{graphicx}
\usepackage[usenames,dvipsnames]{color}
\usepackage{slashed}
\usepackage[colorlinks,citecolor=blue]{hyperref}
\usepackage{pdfpages}
\usepackage{color}
\usepackage{subfig}

\newcommand{\R}{\mbox{\tiny$R$}}
\newcommand{\T}{\mbox{\tiny$T$}}
\newcommand{\s}{\mbox{\tiny$S$}}

\begin{document}

\title{\Large Common Origin of Dirac Neutrino Mass and \\ Freeze-in Massive Particle Dark Matter} 

\author{Debasish Borah}
\email{dborah@iitg.ac.in}
\affiliation{Department of Physics, Indian Institute of Technology
Guwahati, Assam 781039, India}
\author{Biswajit Karmakar}
\email{biswajit@prl.res.in}
\affiliation{Theoretical Physics Division, Physical Research Laboratory, Ahmedabad 380009, India}
\author{Dibyendu Nanda}
\email{dibyendu.nanda@iitg.ac.in}
\affiliation{Department of Physics, Indian Institute of Technology
Guwahati, Assam 781039, India}

\begin{abstract}
Motivated by the fact that the origin of tiny Dirac neutrino masses via the  
standard model Higgs field and non-thermal dark matter populating the Universe 
via freeze-in mechanism require tiny dimensionless couplings of similar order of 
magnitudes $(\sim 10^{-12})$, we propose a framework that can dynamically 
generate such couplings in a unified manner. Adopting a flavour symmetric 
approach based on $A_4$ group, we construct a model where Dirac neutrino 
coupling to the standard model Higgs and dark matter coupling to its mother 
particle occur at dimension six level involving the same flavon fields, thereby 
generating the effective Yukawa coupling of same order of magnitudes. The mother 
particle for dark matter, a complex scalar singlet, gets thermally produced in 
the early Universe through Higgs portal couplings followed by its thermal 
freeze-out and then decay into the dark matter candidates giving rise to the 
freeze-in dark matter scenario.  Some parts of the Higgs portal couplings of the 
mother particle can also be excluded by collider constraints on invisible decay 
rate of the standard model like Higgs boson. We show that the correct neutrino 
oscillation data can be successfully produced in the model which predicts normal 
hierarchical neutrino mass. The model also predicts the atmospheric angle to be 
in the lower octant if the Dirac CP phase lies close to the presently preferred 
maximal value.
\end{abstract}

\maketitle 

\section{Introduction}
Although the non-zero neutrino mass and large leptonic mixing are well  
established facts by now \cite{Patrignani:2016xqp}, with the present status of 
different neutrino parameters being shown in global fit analysis 
\cite{Esteban:2016qun, deSalas:2017kay}, we still do not know a few things about 
neutrinos. They are namely, (a) nature of 
neutrinos: Dirac or Majorana, (b) mass hierarchy of neutrinos: normal $(m_3 > 
m_2 > m_1)$ or inverted $(m_2 > m_1 > m_3)$ and (c) leptonic CP violation as  
well as the octant of atmospheric mixing angle $\theta_{23}$. While neutrino 
oscillation experiments are not sensitive to the nature of neutrinos, 
experiments looking for lepton number violating signatures like neutrinoless 
double beta decay $(0\nu\beta\beta)$ have the potential to confirm the Majorana 
nature of neutrinos with a positive signal. Although the oscillation experiments 
are insensitive to the lightest neutrino mass, the negative results at 
$(0\nu\beta\beta)$ experiments have been able to disfavour the quasi-degenerate 
regime of light neutrino masses. Similarly, the cosmology experiment Planck has 
also constrained the lightest neutrino mass from its bound on the sum of 
absolute neutrino masses $\sum  m_i  \leq 0.17$ eV \cite{Ade:2015xua}.

On the other hand, in cosmic frontier, we have significant amount of  evidences 
\cite{Zwicky:1933gu, Rubin:1970zza, Clowe:2006eq, Hinshaw:2012aka, Ade:2015xua} 
suggesting the presence of non-baryonic form of matter, or the so called Dark 
Matter (DM) in large amount in the present Universe. According to the latest 
cosmology experiment Planck \cite{Ade:2015xua}, almost $26\%$ of the present 
Universe's energy density is in the form of DM while only around $5\%$ is the 
usual baryonic matter leading the rest of the energy budget to mysterious dark 
energy. Quantitatively, the DM abundance at present is quoted as 
$0.1172\leq\Omega_{\rm DM}h^2\leq0.1226$ at 67\% C.L. \cite{Ade:2015xua} where 
$\Omega_{\rm DM} = \rho_{\rm DM}/\rho_{\rm cr}$ is the DM density parameter with 
$\rho_{\rm cr} = \frac{3 H^2_0}{8 \pi G}$ being the critical density of the 
Universe and $H_0$ being the present value of the Hubble parameter. The 
dimensionless parameter $h$ is $H_0/100$. In spite of all these evidences, we do 
not yet know the particle nature of DM.

Since the standard model (SM) of particle physics fails to address the problem  
of neutrino mass and dark matter, several beyond standard model (BSM) proposals 
have been put forward in order to accommodate them. While seesaw mechanism 
\cite{Minkowski:1977sc, GellMann:1980vs, Mohapatra:1979ia, Schechter:1980gr} 
remains the most popular scenario for generating tiny neutrino masses, the 
weakly interacting massive particle (WIMP) paradigm has been the most widely 
studied dark matter scenario. In this framework, a dark matter candidate 
typically with electroweak scale mass and interaction rate similar to 
electroweak interactions can give rise to the correct dark matter relic 
abundance, a remarkable coincidence often referred to as the \textit{WIMP 
Miracle}. Now, if such type of particles whose interactions are of the order of 
electroweak interactions really exist then we should expect their signatures in 
various DM direct detection experiments where the recoil energies of detector 
nuclei scattered by DM particles are being measured. However, after decades of 
running, direct detection experiments are yet to observe any DM-nucleon 
scattering \cite{Tan:2016zwf, Aprile:2017iyp, Akerib:2016vxi}. The absence of 
dark matter signals from the direct detection experiments have progressively 
lowered the exclusion curve in its mass-cross section plane. Although such null 
results could indicate a very constrained region of WIMP parameter space, they 
have also motivated the particle physics community to look for beyond the 
thermal WIMP paradigm where the interaction scale of DM particle can be much 
lower than the scale of weak interaction i.e.\,\,DM may be more 
feebly interacting than the thermal WIMP paradigm. One of the viable 
alternatives of WIMP paradigm, which may
be a possible reason of null results at various direct detection
experiments, is to consider the non-thermal origin of DM \cite{Hall:2009bx}. In this
scenario, the initial number density of DM in the early Universe is
negligible and it is assumed that the interaction strength of DM
with other particles in the thermal bath is so feeble that
it never reaches thermal equilibrium at any epoch in the early Universe. In this set up,
DM is mainly produced from the out of equilibrium decays
of some heavy particles in the plasma. It can also be produced
from the scatterings of bath particles, however if same couplings
are involved in both decay as well as scattering processes
then the former has the dominant contribution to DM relic density
over the latter one \cite{Hall:2009bx, Biswas:2016bfo, Biswas:2017tce}.
The production mechanism for non-thermal DM
is known as freeze-in and the candidates of non-thermal DM produced
via freeze-in are often classified into a group called
Freeze-in (Feebly interacting) massive particle (FIMP). For a recent review of  
this DM paradigm, please see \cite{Bernal:2017kxu}. Similarly, the popular 
seesaw models predict Majorana nature of neutrinos though the results from  
$0\nu \beta \beta$ experiments have so far been negative.  Although such 
negative results do not necessarily prove that the light neutrinos are of Dirac 
nature, it is nevertheless suggestive enough to come up with scenarios 
predicting Dirac neutrinos with correct mass and mixing. This has led to  
several proposals that attempt to generate tiny neutrino masses in a variety of 
ways \cite{Babu:1988yq, Peltoniemi:1992ss, Chulia:2016ngi, Aranda:2013gga, 
Chen:2015jta, Ma:2015mjd, Reig:2016ewy, Wang:2016lve, Wang:2017mcy, Wang:2006jy, 
Gabriel:2006ns, Davidson:2009ha, Davidson:2010sf, Bonilla:2016zef, 
Farzan:2012sa, Bonilla:2016diq, Ma:2016mwh, Ma:2017kgb, Borah:2016lrl, 
Borah:2016zbd, Borah:2016hqn, Borah:2017leo, CentellesChulia:2017koy, 
Bonilla:2017ekt, Memenga:2013vc, Borah:2017dmk, CentellesChulia:2018gwr, 
CentellesChulia:2018bkz, Han:2018zcn}, some of which also accommodate the origin 
of WIMP type dark matter simultaneously.

The present article is motivated by the coincidence that the origin of Dirac 
neutrino masses as well as FIMP dark matter typically require very small 
dimensionless couplings $\sim 10^{-12}$ \cite{Hall:2009bx}. In the neutrino 
sector, such couplings can generate $0.1$ eV Dirac neutrino mass through 
neutrino coupling to the standard model like Higgs. On the other hand, in the 
dark sector, such tiny couplings of the dark matter particle with the mother 
particle makes sure that it gets produced non-thermally through the freeze-in 
mechanism. There have been several attempts where the origin of such feeble 
interactions of DM with the visible sector is generated via higher dimensional 
effective operators \cite{Hall:2009bx, Elahi:2014fsa, McDonald:2015ljz}. Very 
recently, there has been attempt to realise such feeble interactions naturally 
at renormalisable level also \cite{Biswas:2018aib}. The coincidence between such 
tiny FIMP couplings and Dirac neutrino Yukawas was also pointed out, mostly in 
supersymmetric contexts, by the authors of \cite{Hall:2009bx, 
Gopalakrishna:2006kr, deGouvea:2006wd, Page:2007sh, Asaka:2006fs, Asaka:2005cn}. 
Here, we consider an $A_4$ flavour symmetric model\footnote{Similar 
exercise can be carried out using other discrete groups like $A_5$, $S_4$, 
$\Delta(27)$ etc. However, here we adopt $A_4$ flavour symmetry as it is the 
smallest group having a three dimensional representation which in turn helps to 
realise neutrino mixing in an economical way.} where neutrino Dirac mass as well 
as FIMP coupling with its mother particle get generated through dimension six 
operators involving the same flavon fields. A global unbroken lepton number 
symmetry is assumed that forbids the Majorana mass terms of singlet fermions. We 
show that both freeze-in and freeze-out formalisms are important in generating 
the dark matter relic in our scenario. The mother particle, which is long lived 
in this model and decays only to the dark matter at leading order, first freezes 
out and then decays into the dark matter particle. Therefore, the final 
abundance of dark matter particle depends upon the mother particle couplings to 
the standard model particles which can be probed at different ongoing 
experiments. Interestingly, we find that ongoing experiments like the large 
hadron collider (LHC) can probe some part of the parameter space which can give 
rise to sizeable invisible decay of SM like Higgs boson into the long lived 
mother particles. We also show that the correct neutrino oscillation data can be 
reproduced in some specific vacuum alignments of the flavon fields indicating 
the predictive nature of the model. The model also predicts normal hierarchical 
neutrino mass ordering and interesting correlations between neutrino parameters 
requiring the atmospheric mixing angle to be in the lower octant for maximal 
Dirac CP phase.

The remaining part of this letter is organised as follows. In section  
\ref{sec:model} we discuss our $A_4$ flavour symmetric models of Dirac neutrino 
mass and FIMP dark matter and discuss the consequences for neutrino sector for 
some benchmark scenarios. In section \ref{sec:fimp}, we discuss the calculation 
related to relic abundance of dark matter and then finally conclude in section 
\ref{sec:conc}.

\section{$A_4$ Model for Dirac neutrinos and FIMP dark matter}
\label{sec:model}

We first consider a minimal model based on $A_4$ flavour symmetry that can give  
rise to tiny Dirac neutrino masses and mixing at dimension six level. A brief 
details of $A_4$ group is given in appendix \ref{appen1}. The fermion sector of 
the standard model is extended by three copies of gauge singlet right handed 
neutrinos ($\nu_R$) and an additional gauge singlet fermion ($\psi$) which plays 
the role of FIMP dark matter. These right handed neutrinos transform in the same 
way just like the standard model lepton doublets ($L$) do under $A_4$, a typical 
feature of most of the 
$A_4$ flavour symmetric realisations of neutrino mass. We also introduce four  
different flavon fields for the desired phenomenology of neutrino mass and dark 
matter. The $A_4$ flavour symmetry is augmented by additional discrete 
symmetries $Z_4 \times Z^{\prime}_4$ and a global unbroken lepton number 
symmetry $U(1)_L$ in order to forbid the unwanted terms. Transformations of the 
fields under the complete flavour symmetry of the model  $A_4 \times Z_4 \times 
Z^{\prime}_4 \times U(1)_L$ are given in 
Table \ref{tab:a}.  
\begin{table}[h]
\centering
\resizebox{7cm}{!}{%
\begin{tabular}{|c|ccccc|cccc|}
\hline
 Fields & $L$  & $e_{\R}, \mu_{\R}, \tau_{\R}$ &  $H$ & $\nu_{\R}$& $\psi$& 
$\phi_{S}$ & $\phi_{\T}$& $\zeta$ &$\eta$ \\
\hline
$A_{4}$ & 3 & 1,$1''$,$1'$ & 1 & 3&1 & 3 & 3 & 1 &1\\
\hline
$Z_{4}$ & -$i$ &-$i$& 1& $i$ & $i$& $i$&1 & $i$ &1 \\
\hline
 $Z^{\prime}_4$ & 1 & 1 & 1& 1 & $ i$ & 1&1 & $i$ &$-1$\\
 \hline
 $U(1)_L$ & 1 & 1 & 0 & 1 & 0 & 0 & 0 & 0 & 0 \\
 \hline
\end{tabular}
}\
\caption{\label{tab:a} Field content and transformation properties under
$A_4 \times Z_4 \times Z^{\prime}_4 \times U(1)_L $ symmetry of the model. }
\end{table}
The construction here includes two $A_4$ triplet flavons, $\phi_{\T}$ and   
$\phi_{\s}$, which play a crucial role in generating masses and mixing for charged 
leptons and Dirac neutrinos respectively. Now, 
for charged lepton sector, the relevant Yukawa Lagrangian can be written as 
\begin{equation}\label{Lag:cl2}
 \mathcal{L}_l =  \frac{y_e}{\Lambda}(\bar{L}\phi_{\T})H e_{\R}
+\frac{y_{\mu}}{\Lambda}(\bar{L}\phi_{\T})_{1'}H\mu_{\R}+ 
\frac{y_{\tau}}{\Lambda}(\bar{L}\phi_{\T})_{1''}H\tau_{\R}
\end{equation}
where $\Lambda$ is the cut-off scale of the theory. Here and subsequently 
all the $y$'s stand for the respective coupling constants, unless otherwise 
mentioned. The leading contributions to the charged lepton mass via $\bar{L}H 
\ell_i$ 
(where $\ell_i$ are the RH charged leptons) are not allowed due to the specific 
$A_4$ symmetry. When the triplet flavon $\phi_{\T}$ is present in the model it leads to an $A_4$ 
invariant dimension five operator as given in equation (\ref{Lag:cl2}) which subsequently generates the relevant masses after flavons and the SM Higgs field acquire non-zero vacuum expectation value (vev)'s. Using the $A_4$ product rules given in appendix \ref{appen1} and taking generic triplet flavon vev 
alignment $\langle \phi_T \rangle=(v_T, v_T, v_T)$, we can write down the charged lepton mass matrix as
\begin{eqnarray}\label{mCL2}
m_{l} =\frac{vv_{\T}}{\Lambda} \left(
\begin{array}{ccc}
         y_e & y_{\mu}          & y_{\tau}\\
         y_e & \omega y_{\mu}   & \omega^2 y_{\tau} \\
         y_e & \omega^2 y_{\mu} & \omega y_{\tau} 
\end{array}
\right).
\end{eqnarray}
Here $v$ denotes the vev of the SM Higgs doublet $H$ and 
$\omega=e^{i2\pi/3}$ is the cube root of unity. This mass matrix can be 
diagonalised by 
using the magic matrix $U_{\omega}$, given by
\begin{eqnarray}\label{eq:omega}
U_{\omega} =\frac{1}{\sqrt{3}}\left(
\begin{array}{ccc}
         1 & 1          & 1\\
         1 & \omega     & \omega^2\\
         1 & \omega^2   & \omega 
\end{array}
\right).  
\end{eqnarray}
Now, as indicated earlier, the complete $A_4 \times Z_4 \times Z^{\prime}_4$ discrete 
symmetry plays an instrumental role in generating tiny Dirac neutrino mass and mixing at dimension six level. Any contribution to the neutrino mass (through $\bar{L}\tilde{H}\nu_R$) is 
forbidden up to dimension five level in the present set-up. Since charged lepton masses are generated at dimension five level, it naturally explains the observed hierarchy between charged and neutral lepton masses. Presence of the $A_4$ triplet flavon $\phi_{S}$ generates the required dimension six operator for neutrino mass and mixing. The relevant Yukawa Lagrangian for neutrino sector is given by 
\begin{eqnarray}\label{modela}
 \mathcal{L}_{\nu}&=&\bar{L}\tilde{H}\nu_R 
\frac{(\phi_S)^2}{\Lambda^2}+\text{h.c.}\\
            &=&\frac{y_{s}}{\Lambda^2}(\bar{L}\tilde{H}\nu_R)_S (\phi_S 
\phi_S)_S+\frac{y_{s'}}{\Lambda^2}(\bar{L}\tilde{H}\nu_R)_S (\phi_S 
\phi_S)_A+\frac{y_{a}}{\Lambda^2}(\bar{L}\tilde{H}\nu_R)_A (\phi_S 
\phi_S)_S\nonumber\\
&+&\frac{y_{a'}}{\Lambda^2}(\bar{L}\tilde{H}\nu_R)_A (\phi_S 
\phi_S)_A+\frac{y_{x_1}}{\Lambda^2}(\bar{L}\tilde{H}\nu_R)_1 (\phi_S 
\phi_S)_1+\frac{y_{x_2}}{\Lambda^2}(\bar{L}\tilde{H}\nu_R)_{1'} (\phi_S 
\phi_S)_{1''}\nonumber\\
&+&\frac{y_{x_3}}{\Lambda^2}(\bar{L}\tilde{H}\nu_R)_{1''} (\phi_S 
\phi_S)_{1'}+\text{h.c.}.
\end{eqnarray}
Here the subscripts $S$ and $A$ stands for symmetric and anti-symmetric parts 
of $A_4$ triplets products (see Appendix \ref{appen1} for details) in the $S$ 
diagonal basis adopted in the analysis and $1, 1'$ and $1''$ stand for three 
singlets of $A_4$. For the most general vev alignment $\langle \phi_S 
\rangle=(v_{S_1},v_{S_2},v_{S_3})$, the effective mass matrix for neutrinos can 
be written as 
\begin{eqnarray}\label{mnu:gen}
m_{\nu}&=&\left(
\begin{array}{ccc}
 x_{11} & s_{21} + a_{21} & s_{31}+a_{31} \\
 s_{21} - a_{21} & x_{22} & s_{32} + a_{32}\\
 s_{31} - a_{31} & s_{32} - a_{32} & x_{33}
\end{array}
\right)
\end{eqnarray}
where the diagonal elements are given by
\begin{footnotesize}
\begin{align}\label{xs}
x_{11}=&\left[y_{x_1}(v_{S_1}^2+v_{S_2}^2+v_{S_3}^2)+y_{x_2}(v_{S_1}
^2+\omega^2 v_{S_2} ^2+\omega v_{ S_3 } ^2)+y_{x_3}(v_ { S_1 
}^2+\omega v_{S_2}^2+\omega^2 v_{S_3}^2)\right]v/\Lambda^2,\\
x_{22}=&\left[y_{x_1}(v_{S_1}^2+v_{S_2}^2+v_{S_3}^2)+y_{x_2}\omega(v_{S_1}
^2+\omega^2 v_{S_2} ^2+\omega v_{ S_3 } ^2)+y_{x_3}\omega^2(v_ { S_1 
}^2+\omega v_{S_2}^2+\omega^2 v_{S_3}^2)\right]v/\Lambda^2,\\
x_{33}=&\left[y_{x_1}(v_{S_1}^2+v_{S_2}^2+v_{S_3}^2)+y_{x_2}\omega^2(v_{S_1}
^2+\omega^2 v_{S_2} ^2+\omega v_{ S_3 } ^2)+y_{x_3}\omega(v_ { S_1 
}^2+\omega v_{S_2}^2+\omega^2 v_{S_3}^2)\right]v/\Lambda^2.
\end{align}
\end{footnotesize}
Now, the symmetric part originated from $A_4$ triplet products are given by 
\begin{footnotesize}
\begin{align}\label{ss}
 s_{32}=&y_sv(v_{S_2}v_{S_3}+v_{S_3}v_{S_2})/\Lambda^2+ 
y_s'v(v_{S_2}v_{S_3}-v_{S_3} v_{S_2})/\Lambda^2,\\
 s_{31}=&y_sv(v_{S_3}v_{S_1}+v_{S_1}v_{S_3})/\Lambda^2+ 
y_s'v(v_{S_3}v_{S_1}-v_{S_1} v_{S_3})/\Lambda^2,\\
s_{21}=&y_sv(v_{S_1}v_{S_2}+v_{S_2}v_{S_1})/\Lambda^2+ 
y_s'v(v_{S_1}v_{S_2}-v_{S_2} v_{S_1})/\Lambda^2.
\end{align}
\end{footnotesize}
As seen above, when neutrinos are Dirac fermions instead of Majorana, then there 
 is an additional anti-symmetric contribution in the neutrino mass 
matrix which remains absent in the Majorana case due to symmetric property of  
the Majorana mass term. This additional contribution can in fact explain nonzero 
$\theta_{13}$ in a more economical setup~\cite{Memenga:2013vc,Borah:2017dmk, 
Borah:2017qdu} compared to the one for 
Majorana neutrinos~\cite{Karmakar:2014dva}. In the mass matrix given by 
equation (\ref{mnu:gen}) these anti-symmetric contributions are given by
\begin{footnotesize}
\begin{align}\label{as}
a_{32}=&y_av(v_{S_2}v_{S_3}+v_{S_3}v_{S_2})/\Lambda^2+ 
y_a'v(v_{S_2}v_{S_3}-v_{S_3} v_{S_2})/\Lambda^2,\\
 a_{31}=&y_av(v_{S_3}v_{S_1}+v_{S_1}v_{S_3})/\Lambda^2+ 
y_a'v(v_{S_3}v_{S_1}-v_{S_1} v_{S_3})/\Lambda^2,\\
a_{21}=&y_sv(v_{S_1}v_{S_2}+v_{S_2}v_{S_1})/\Lambda^2+ 
y_a'v(v_{S_1}v_{S_2}-v_{S_2} v_{S_1})/\Lambda^2.
\end{align}
\end{footnotesize}
The most general mass matrix for Dirac neutrinos given in equation (\ref{mnu:gen}) 
can be further simplified depending upon the specific and simpler vev alignments of the triplet flavon 
$\phi_{S}$. Here we briefly discuss a few such possible alignments analytically and then 
restrict ourselves to one such scenario for numerical analysis which can explain neutrino masses 
and mixing in a minimal way. Note that such vev alignments demand a complete analysis of the scalar sector of the model and can be obtained in principle, from the minimisation of the scalar potential~\cite{He:2006dk, Branco:2009by, Lin:2008aj,Dorame:2012zv, 
Rodejohann:2015hka}. For simplicity, when we consider the vev alignment of $\phi_S$ to be 
$\langle \phi_S \rangle=(v_{S},v_{S},v_{S})$ from equation (\ref{xs})-(\ref{as}), we 
obtain $s_{32}=s_{31}=s_{21}=2y_svv_{S}^2/\Lambda^2=s$ (say) and 
$a_{32}=a_{31}=a_{21}=2y_avv_{S}^2/\Lambda^2=a$ (say). Hence the neutrino mass matrix 
takes the form 
\begin{eqnarray}\label{mix1}
m_{\nu}&=&\left(
\begin{array}{ccc}
 x & s+a & s+a \\
 s-a & x & s+a\\
 s-a & s-a & x
\end{array}
\right),
\end{eqnarray}
where $x_{11}=x_{22}=x_{33}=3y_{x_1}vv_{S}^2/\Lambda^2=x$ (say). For even 
more simplified\footnote{Vev alignments like  $\langle \phi_S 
\rangle=(0,0,v_{S}), (0,v_{S},0)$ and $(v_{S},0,0)$ are not allowed in the 
present construction of Dirac neutrino mass.} scenarios of vev alignments 
$\langle \phi_S\rangle=(v_{S},v_{S},0)$ and 
$\langle \phi_S \rangle=(0,v_{S},v_{S})$ the neutrino mass matrices are given 
by \begin{eqnarray}\label{mix24}
m_{\nu}&=&\left(
\begin{array}{ccc}
 x_{11} & s+a & 0 \\
 s-a & x_{22} & 0\\
 0 & 0 & x_{33}
\end{array}
\right);~~~~ m_{\nu}=\left(
\begin{array}{ccc}
 x_{11} & 0 & 0 \\
 0 & x_{22} & s+a\\
 0 & s-a & x_{33}
\end{array}
\right), 
\end{eqnarray} 
respectively, where the elements are defined as $s_{21}=2y_svv_{S}^2/\Lambda^2=s$ (say), 
$a_{21}=2y_avv_{S}^2/\Lambda^2=a$ (say), 
$x_{11}=(2y_{x_1}-y_{x_2}\omega-y_{x_3}\omega^2)vv_{S}^2/\Lambda^2$, 
$x_{22}=(2y_{x_1}-y_{x_2}\omega^2-y_{x_3}\omega)vv_{S}^2/\Lambda^2$, 
$x_{33}=(2y_{x_1}-y_{x_2}-y_{x_3})vv_{S}^2/\Lambda^2$ and 
$s_{32}=2y_svv_{S}^2/\Lambda^2=s$ (say),
$a_{32}=2y_avv_{S}^2/\Lambda^2=a$ (say), 
$x_{11}=(2y_{x_1}-y_{x_2}-y_{x_3})vv_{S}^2/\Lambda^2$, 
$x_{22}=(2y_{x_1}-y_{x_2}\omega-y_{x_3}\omega^2)vv_{S}^2/\Lambda^2$,  
$x_{33}=(2y_{x_1}-y_{x_2}\omega^2-y_{x_3}\omega)vv_{S}^2/\Lambda^2$ 
respectively. As evident from these two neutrino mass matrices given by equation 
(\ref{mix24}), a Hermitian matrix ($m_{\nu}m_{\nu}^{\dagger}$) obtained from 
these demands a rotation in the 12 and 23 planes respectively. This, however,  
is not sufficient to to explain observed neutrino mixing along with the 
contribution ($U_{\omega}$) from the charged lepton sector given in equation 
(\ref{eq:omega}). Now, a third possibility with vev alignment $\langle \phi_S 
\rangle=(v_{S},0,v_{S})$, yields a compatible neutrino mass matrix, 
given by,
\begin{eqnarray}\label{mix3}
m_{\nu}&=&\left(
\begin{array}{ccc}
 x_{11} & 0 & s+a \\
 0 & x_{22} & 0\\
 s-a & 0 & x_{33}
\end{array}
\right), 
\end{eqnarray}
where $s_{31}=2y_svv_{S}^2/\Lambda^2=s$ (say),
$a_{31}=2y_avv_{S}^2/\Lambda^2=a$ (say),   
$x_{11}=(2y_{x_1}-y_{x_2}\omega^2-y_{x_3}\omega)vv_{S}^2/\Lambda^2$, 
$x_{22}=(2y_{x_1}-y_{x_2}-y_{x_3})vv_{S}^2/\Lambda^2$ and
$x_{33}=(2y_{x_1}-y_{x_2}\omega-y_{x_3}\omega^2)vv_{S}^2/\Lambda^2$. Although 
parameters present here are in general complex, for the diagonal elements we consider them 
to be equal that is, $x_{11}=x_{22}=x_{33}=x$ and real without loss of any 
generality. Now, to diagonalise this mass matrix, let us first define a Hermitian 
matrix as 
\begin{eqnarray}\label{mmd}
 \mathcal{M}&=&m_{\nu}m_{\nu}^{\dagger}\nonumber\\
            &=&\left(
 \begin{array}{ccc}
 x^2+|s + a|^2 & 0   & x(s - a)^*+x(s+a) \\
 o& x^2 & 0\\
 x(s - a)+x(s+a)^*& 0 & x^2+|s - a|^2
\end{array}
\right). 
\end{eqnarray}
Here the complex terms corresponding to the symmetric and anti-symmetric parts of $A_4$ products can be written as $s=|s|^{i\phi_{s}}$ and $a=|a|^{i\phi_{a}}$. These complex phases essentially 
dictates the CP violation of the theory. Clearly, the structure of 
$\mathcal{M}$ given in equation (\ref{mmd}) indicates rotation in the 13 plane 
through the relation $U_{13}^{\dagger}\mathcal{M}U_{13}={\rm 
diag}(m_1^2,m_2^2,m_3^2)$ is sufficient to diagonalise this matrix, where the 
$U_{13}$ is given by 
\begin{eqnarray}\label{u13}
U_{13}=\left(
\begin{array}{ccc}
 \cos\theta               & 0 & \sin\theta{e^{-i\psi}} \\
     0                    & 1 &            0 \\
 -\sin\theta{e^{i\psi}} & 0 &        \cos\theta
\end{array}
\right),  
\end{eqnarray}
\begin{figure}[h]
$$
\includegraphics[height=5cm]{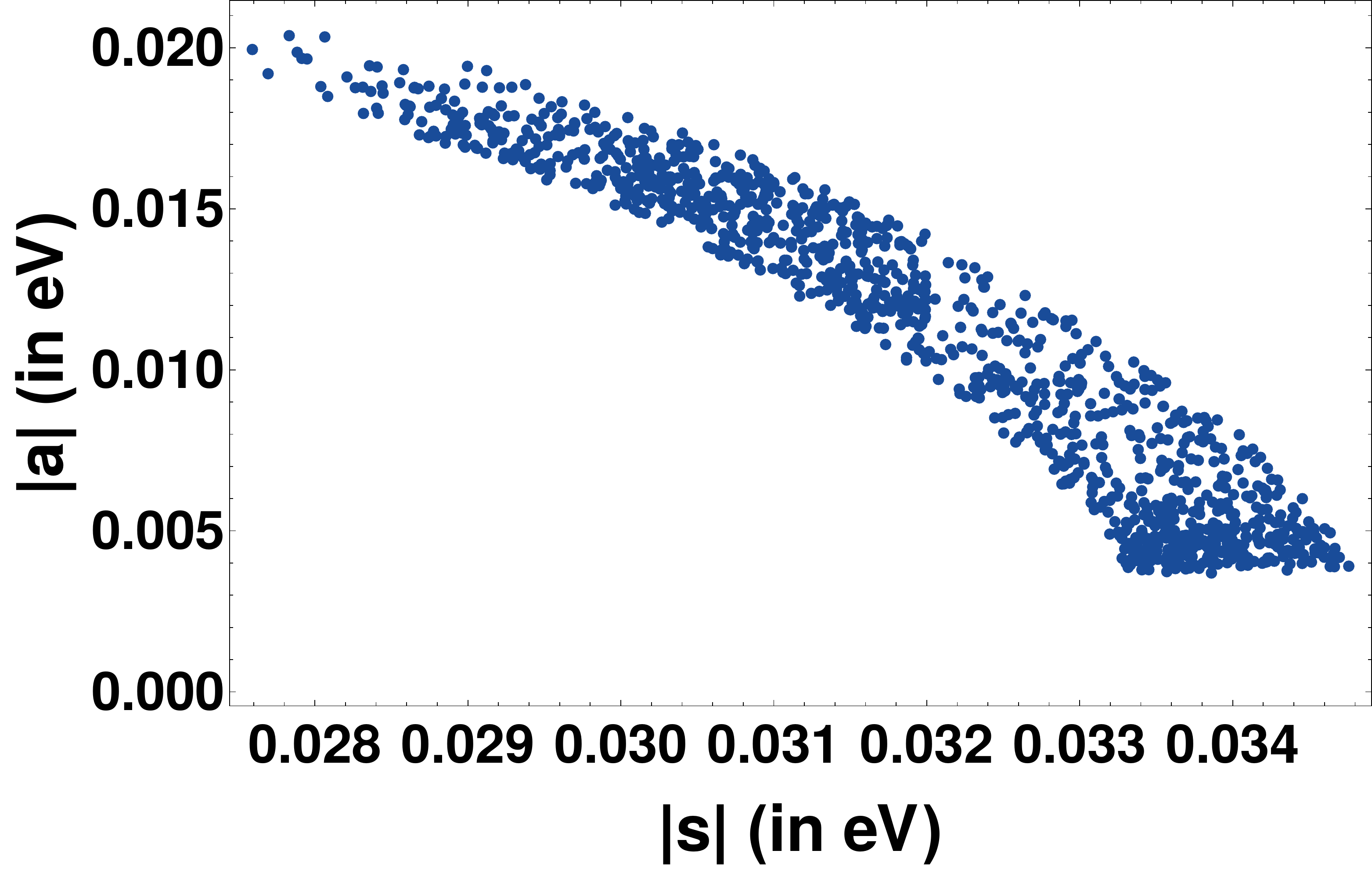}~
\includegraphics[height=5.1cm]{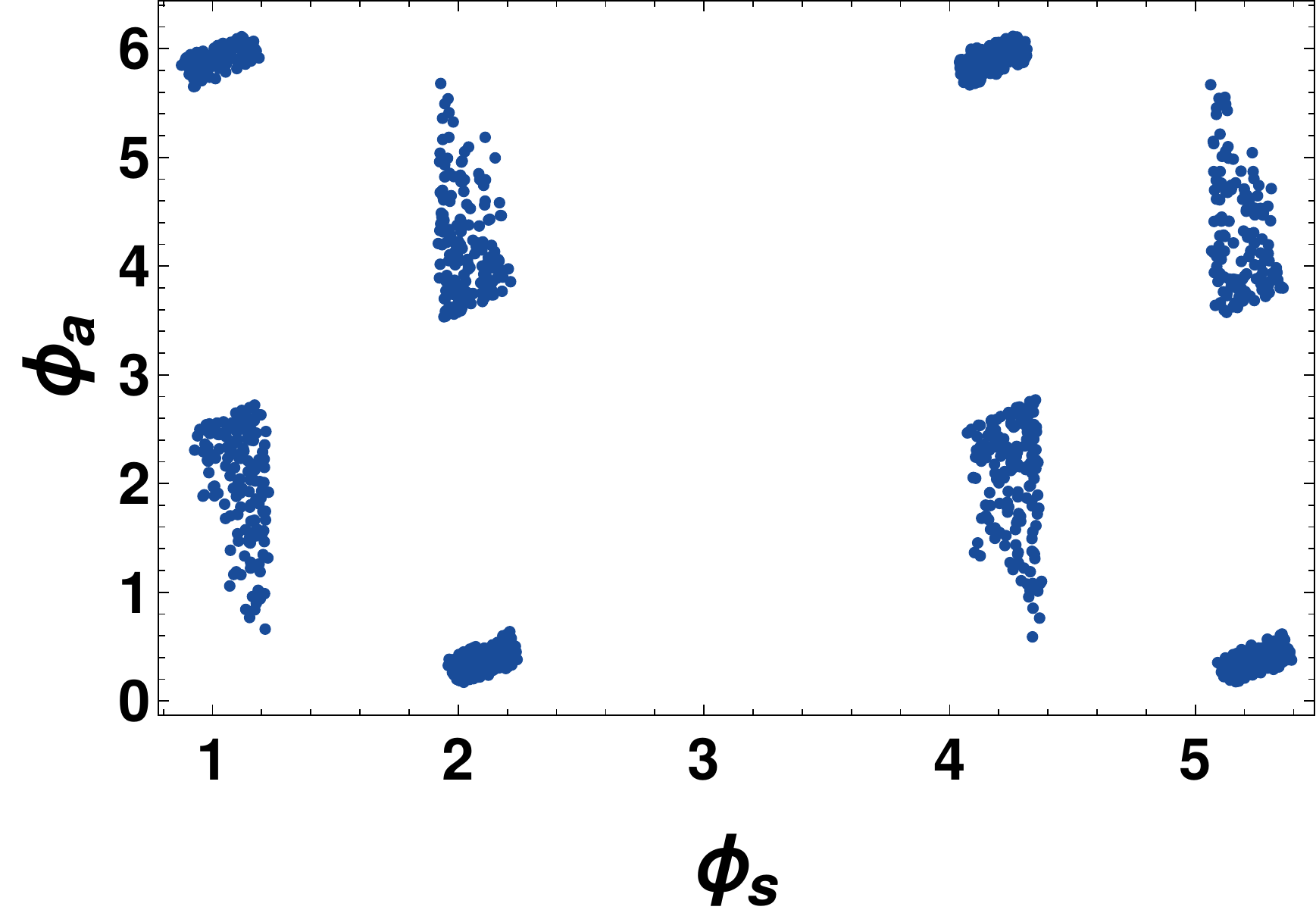}
$$
\caption{Correlations between different model parameters for 3$\sigma$ 
allowed ranges of $\theta_{13}$, $\theta_{12}$, $\theta_{23}$, mass squared 
differences~\cite{Esteban:2016qun} and $\sum m_i \leq 0.17$ 
eV. Left panel: the allowed points in $|s|-|a|$ 
(symmetric and anti-symmetric parts of the neutrino mass matrix respectively) 
plane. Right panel: the allowed points in the 
$\phi_{s}-\phi_{a}$ (phases associated with symmetric  nd anti-symmetric parts 
respectively) plane.}
\label{fig:as}
\end{figure}
and the mass eigenvalues are found to be 
\begin{eqnarray}
 m_1^2&=&x^2+A-B,\label{eq:tm1}\\
 m_2^2&=&x^2,\label{eq:tm2}\\
 m_3^2&=&x^2+A+B,\label{eq:tm3}
\end{eqnarray}
 where $A=|s|^2+|a|^2$ and $B=\sqrt{(2|s||a|\cos(\phi_{s}-\phi_{a}
))^2+4x^2(|s|^2\cos^2\phi_{s}+|a|^2\sin^2\phi_{a}) }$. One important inference 
of such ordering is that inverted hierarchy of neutrino mass is not feasible in this setup
as $\Delta m^2_{23}+\Delta m^2_{21}=-2(|s|^2+|a|^2)<0$, implying $m_3>m_2$. Also, the two 
parameters $\theta$ and $\psi$ appearing in $U_{13}$ can be expressed as
\begin{eqnarray}\label{eq:ang}
 \tan 2\theta=\frac{x(|a|\sin\phi_{a}\sin\psi-|s|\cos\phi_{s}\cos\psi)}
 {|s||a|\cos(\phi_{s}-\phi_{a})}~~~ {\rm and}~~~ 
\tan\psi=-\frac{|a|\sin\phi_{a}}
 {|s|\cos\phi_{s}},
\end{eqnarray}  
in terms of the parameters appearing in the mass matrix. Hence the final lepton 
mixing matrix is given by 
\begin{eqnarray}\label{uour}
 U&=&U^{\dagger}_{\omega}U_{13}. 
\end{eqnarray}
Comparing this with the Pontecorvo Maki Nakagawa Sakata (PMNS) mixing matrix parametrised as
\begin{equation}
U_{\text{PMNS}}=\left(\begin{array}{ccc}
c_{12}c_{13}& s_{12}c_{13}& s_{13}e^{-i\delta}\\
-s_{12}c_{23}-c_{12}s_{23}s_{13}e^{i\delta}& 
c_{12}c_{23}-s_{12}s_{23}s_{13}e^{i\delta} & s_{23}c_{13} \\
s_{12}s_{23}-c_{12}c_{23}s_{13}e^{i\delta} & 
-c_{12}s_{23}-s_{12}c_{23}s_{13}e^{i\delta}& c_{23}c_{13}
\end{array}\right),  
\label{PMNS}
\end{equation}
one can obtain correlations between neutrino mixing angles $\theta_{13}, 
\theta_{12}, \theta_{23}$, Dirac CP phase $\delta$ and parameters appearing in 
equation (\ref{uour}) very easily~\cite{Memenga:2013vc, Grimus:2008tt, Albright:2008rp, 
Albright:2010ap, He:2011gb, Borah:2017dmk}. Hence, from equations 
(\ref{eq:ang}-\ref{PMNS}) it is evident that  the mixing angles ($\theta_{13}, 
\theta_{12}, \theta_{23}$) and Dirac CP phase $(\delta)$ involved in the lepton 
mixing matrix $U_{\text{PMNS}}$ are functions of $x$, $|s|$, $|a|$, $\phi_{s}$ 
and $\phi_{a}$. Neutrino mass eigenvalues are also function of these parameters 
as obtained in equations (\ref{eq:tm1}-\ref{eq:tm3}). These parameters can 
be constrained using the current data on neutrino mixing angles and mass 
squared differences~\cite{Esteban:2016qun,deSalas:2017kay}. Here in our 
analysis we adopt the 3$\sigma$ variation of neutrino oscillation data obtained 
from the global fit~\cite{Esteban:2016qun} to do so. In figure \ref{fig:as} we have plotted the allowed parameter 
values in $|s|-|a|$ plane (left panel) and $\phi_{s}-\phi_a$ plane (right 
panel) respectively satisfying 3$\sigma$ range of neutrino mixing data as mentioned earlier. 

\begin{figure}[h]
$$
\includegraphics[height=5cm]{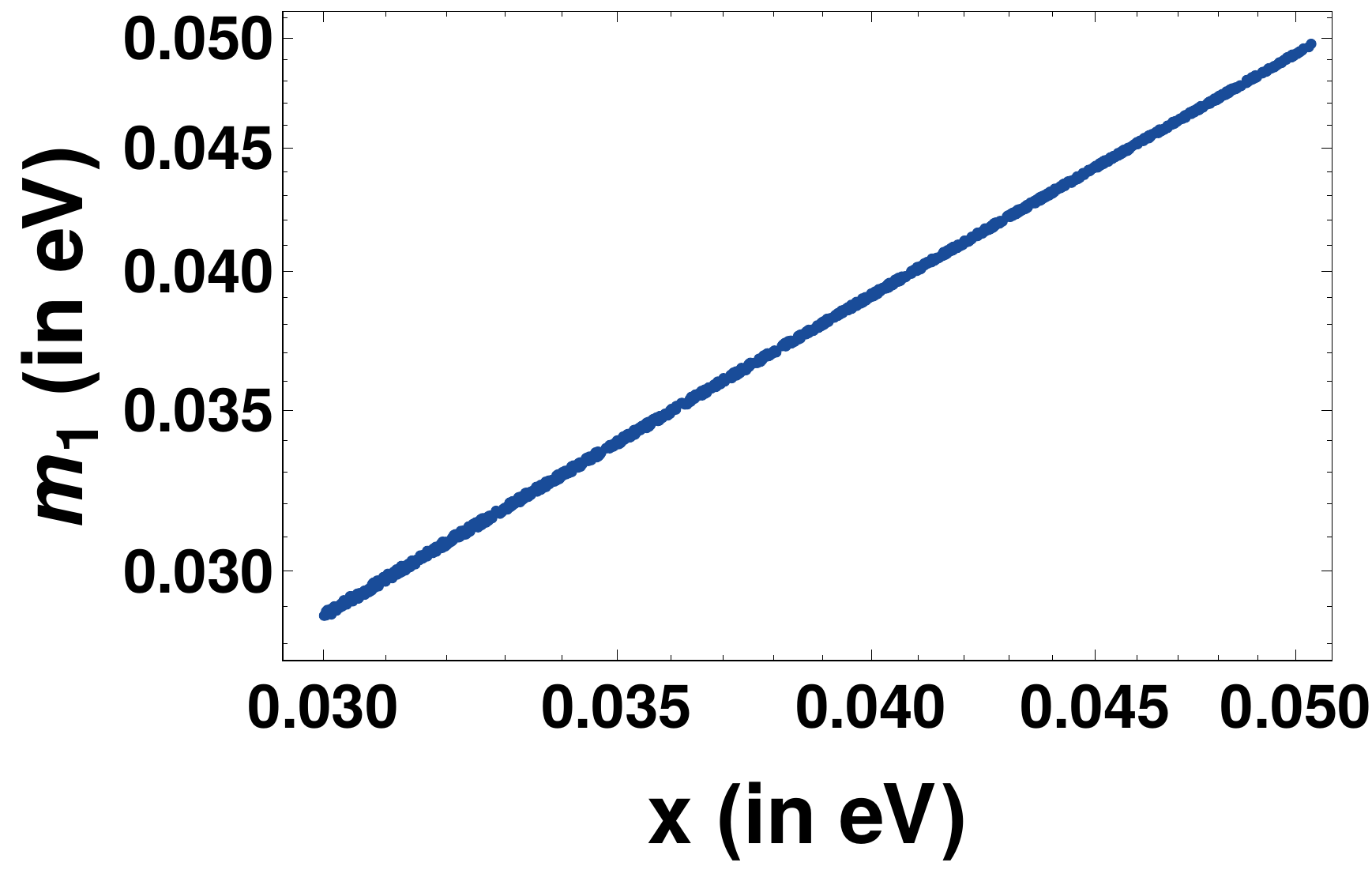}~~
\includegraphics[height=5cm]{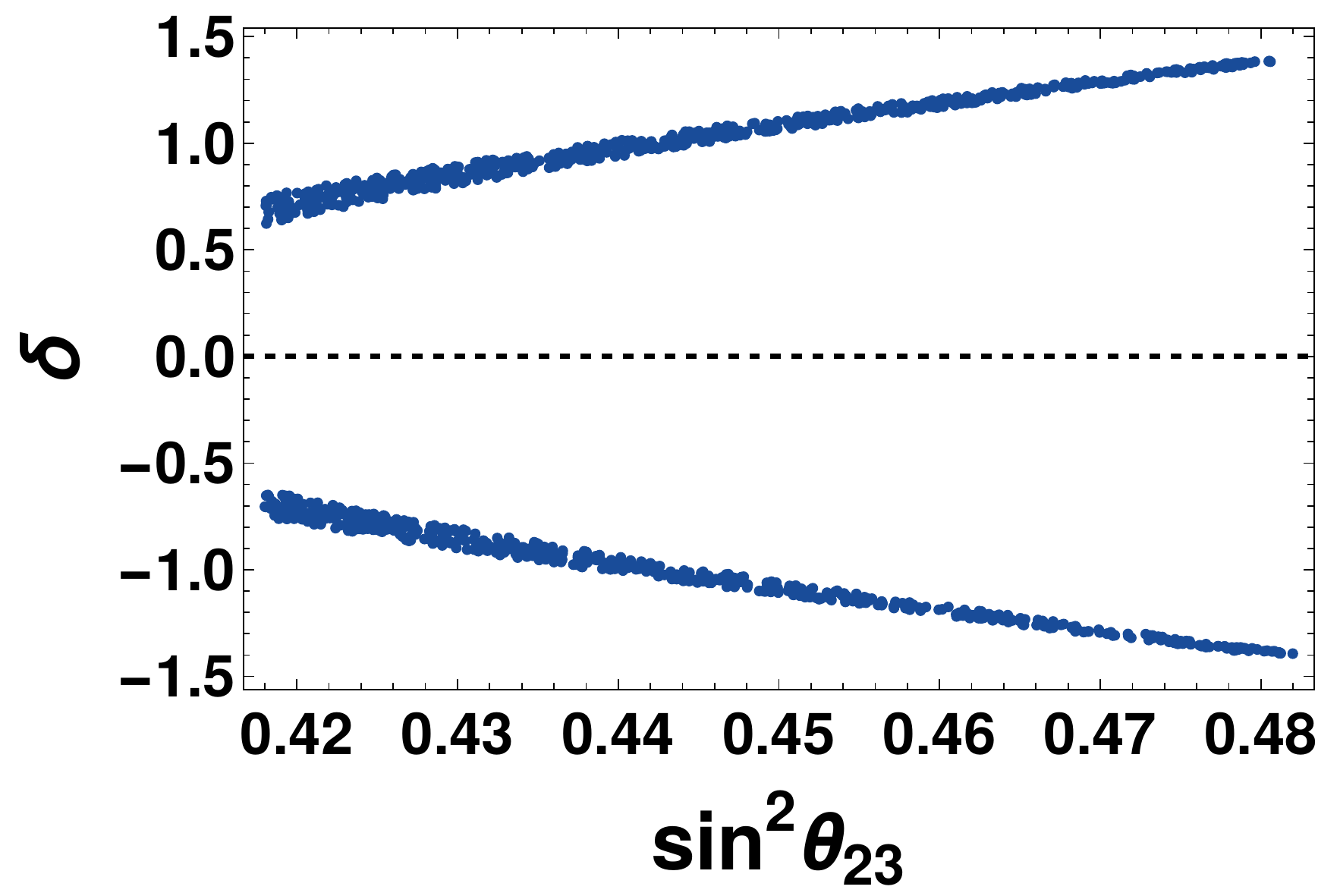}
$$
\caption{Correlations between different light neutrino as well as model parameters for 3$\sigma$ allowed ranges of $\theta_{13}$, $\theta_{12}$, $\theta_{23}$, mass squared 
differences~\cite{Esteban:2016qun} and $\sum m_i \leq 0.17$ eV. Left panel: the lightest neutrino mass is shown as a function of model parameter $x$. Right panel: the predicted correlation between Dirac CP phase $\delta$ and atmospheric mixing angle $\theta_{23}$. }
\label{fig:mplot}
\end{figure}

After constraining the model parameters, the predictions for absolute neutrino mass ($m_1$ in case of normal hierarchy) is plotted in the left panel of figure \ref{fig:mplot}. In this figure, the lightest 
neutrino mass ($m_1$) is shown as a function of the diagonal element of neutrino 
mass matrix $x$. Whereas in the right panel of figure \ref{fig:mplot} we present the correlation 
between Dirac CP phase $\delta$ and $\sin^2\theta_{23}$. Interestingly the 
model predicts $\delta$ in the range $-\pi/2\lesssim \delta \lesssim -\pi/5$ 
and $\pi/5\lesssim \delta \lesssim \pi/2$ whereas $\sin^2\theta_{23}$ lies  in 
the lower octant. Here it is worth mentioning that, the presently preferred value $\delta \sim \pm \pi/2$ as indicated in global fit analysis \cite{Esteban:2016qun}, predicts the atmospheric mixing angle $\theta_{23}$ to be in the lower octant within our framework, as seen from the right panel of figure \ref{fig:mplot}. 

\noindent
{\bf FIMP interactions}: After studying the neutrino sector, we briefly comment upon the Yukawa Lagrangian involving the FIMP dark matter candidate $\psi$ upto dimension six level. From the field content shown in Table \ref{tab:a}, it is obvious that a bare mass term for $\psi$ is not allowed. However, we can generate its mass at dimension five level (same as that of charged leptons). The corresponding Yukawa Lagrangian is 
\begin{equation}
\mathcal{L}_{\psi \zeta} = \frac{1}{2}Y_{\psi \zeta} \frac{\zeta^2}{\Lambda} \psi \psi.
\label{fimpmass}
\end{equation}
Once $\zeta$ acquires a non-zero vev, we can generate a mass $M_{\psi} = Y_{\psi \zeta} \frac{\langle \zeta \rangle ^2}{\Lambda}$. Another important Yukawa interaction of $\psi$ is with the singlet flavon $\eta$ that arises at dimension six level, given by
\begin{eqnarray}
 \mathcal{L}_{\psi \eta}&=&\frac{(\phi_S)^2}{\Lambda^2} \eta\psi\psi\\
                   &=& \frac{v_{S_1}^2+v_{S_2}^2+v_{S_3}^2} {\Lambda^2} \eta
\psi\psi~~{\text{for}}~~\langle \phi_S \rangle=(v_{S_1},v_{S_2},v_{S_3})\\ 
                   &=& \frac{3 v_S^2}{\Lambda^2}\eta\psi\psi 
~~{\text{for}}~~\langle \phi_S \rangle=(v_{S},v_{S},v_{S})\\
&=& \frac{2 v_S^2}{\Lambda^2}\eta\psi\psi 
~~{\text{for}}~~\langle \phi_S \rangle=(v_{S},v_{S},0), (v_{S},0,v_{S}),~{\rm 
or}~(0,v_{S},v_{S})
\end{eqnarray}
It is interesting to note that the same flavon field $\phi_S$ and the ratio $\frac{ \langle \phi_S \rangle^2}{\Lambda^2}$ generates the effective coupling of $\eta-\psi-\psi$ as well as $H-\nu_L-\nu_R$ as discussed earlier in equation \eqref{modela}. We will use these interactions while discussing the dark matter phenomenology in the next section.

\section{Freeze-in Dark Matter}
\label{sec:fimp}
In this section, we discuss the details of calculation related to the relic abundance of FIMP dark matter candidate $\psi$. As per requirement for such dark matter \cite{Hall:2009bx}, the interactions of dark matter particle with the visible sector ones are so feeble that it never attains thermal equilibrium in the early Universe. In the simplest possible scenario of this type, the dark matter candidate has negligible initial thermal abundance and gets populated later due to the decay of a mother particle. Such non-thermal dark matter scenario which gets populated in the Universe through freeze-in (rather than freeze-out of WIMP type scenarios) should have typical coupling of the order $10^{-12}$ with the decaying mother particle. Unless such decays of mother particles into dark matter are kinematically forbidden, the contributions of scattering to freeze-in of dark matter remains typically suppressed compared to the former.

\begin{figure}[h!!!!!!]
\centering
\begin{tabular}{cc}
\includegraphics[width=0.5\textwidth]{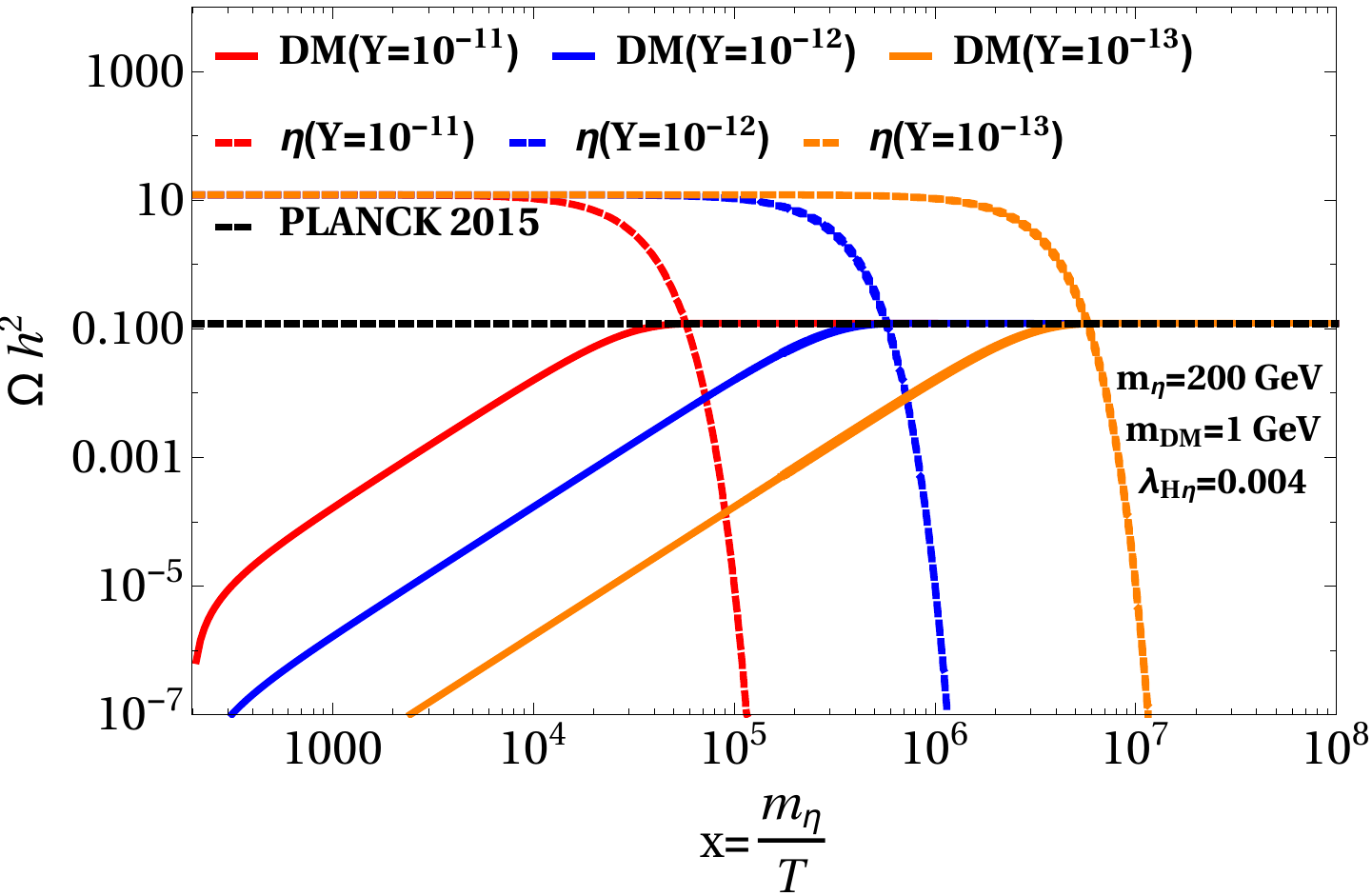}
\includegraphics[width=0.5\textwidth]{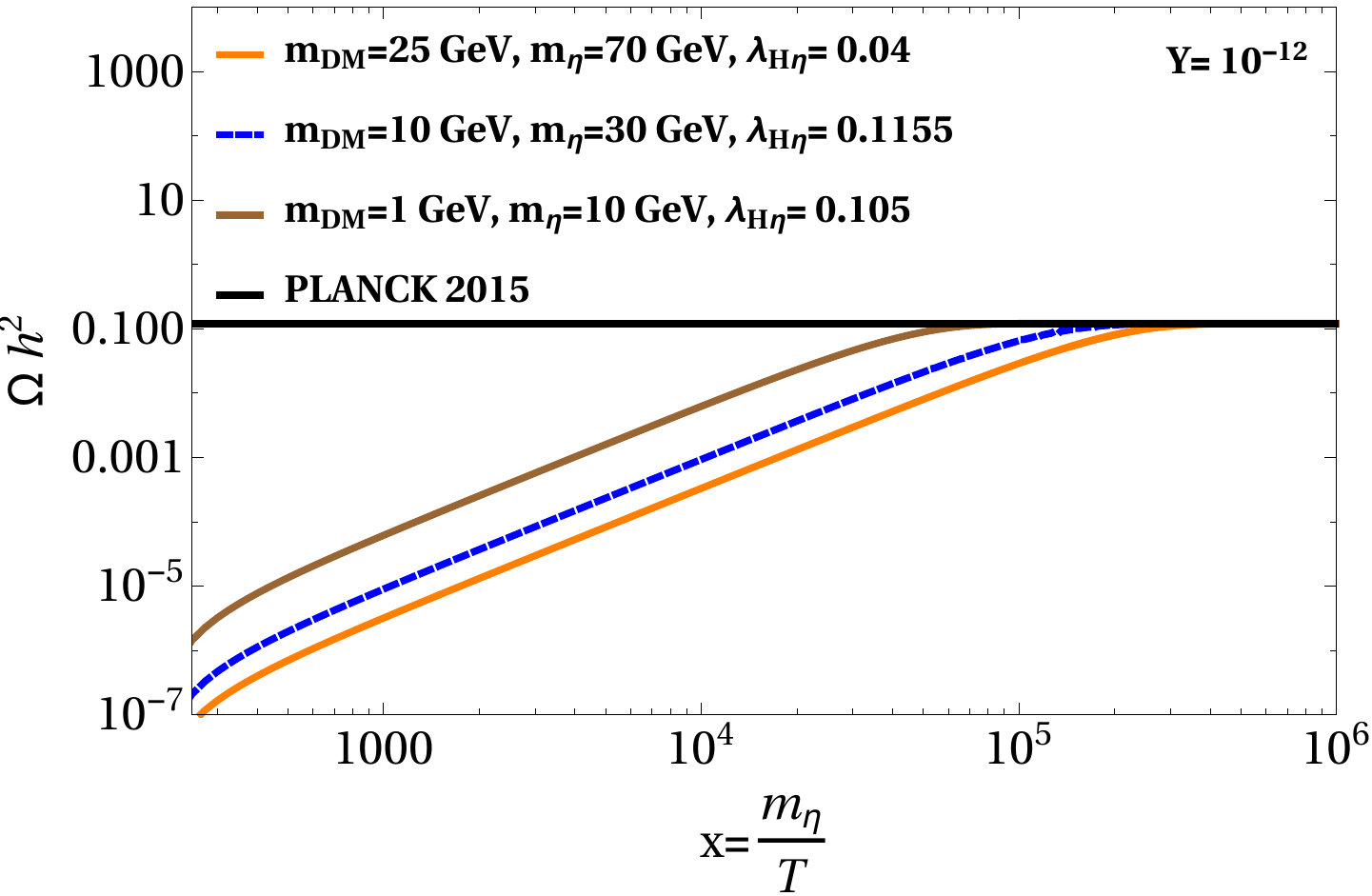}
\end{tabular}
\caption{Left panel: Relic abundance of $\eta$ and dark matter $(\psi)$ as a function of temperature for different benchmark values of Yukawa coupling $Y$. Right panel: Relic abundance of dark matter $(\psi)$ as a function of temperature for different benchmark values of $\psi$ and $\eta$ masses and fixed Yukawa $Y=10^{-12}$.}
\label{fig1}
\end{figure}

In our model, the fermion $\psi$ naturally satisfies the criteria for being a FIMP dark matter candidate without requiring highly fine-tuned couplings mentioned above. This is due to the fact that this fermion is a gauge singlet and its leading order interaction to the mother particle $\eta$ arises only at dimension six level. As discussed in the previous section, the effective Yukawa coupling for $\eta \psi \psi$ interaction is dynamically generated by flavon vev's $Y \sim \frac{v^2_S}{\Lambda^2}$. Now, the decay width of $\eta$ into two dark matter particles $(\psi)$ can be written as 
\begin{equation}
\Gamma_{\eta\rightarrow \psi \psi} = \frac{Y^2 \left(m_\eta ^2-4 m_\psi^2\right) \sqrt{1-\frac{4 m_\psi ^2}{m_\eta ^2}}}{8 \pi m_\eta }
\end{equation}
where Y is the effective Yukawa coupling, m$_\eta$ and m$_\psi$ are the masses of the mother particle and $\psi$ respectively. From the transformation of the singlet scalar $\eta$ under the symmetry group of the model, it is clear that it does not have any linear term in the scalar sector and hence does not have any other decay modes apart from the one into two dark matter particles. Since this decay is governed by a tiny effective Yukawa coupling, this makes the singlet scalar long lived. However, this singlet scalar can have sizeable quartic interactions with other scalars like the standard model Higgs doublet and hence can be thermally produced in the early Universe. Now, considering the mother particle $\eta$ to be in thermal equilibrium in the early Universe which also decays into the dark matter particle $\psi$, we can write down the relevant Boltzmann equations for co-moving number densities of $\eta, \psi$ as
\begin{eqnarray}
\frac{dY_{\eta}}{dx} &=& -\frac{4 \pi^2}{45}\frac{M_{\rm Pl} M_{\rm sc}}{1.66}\frac{\sqrt{g_{\star}(x)}}{x^2}\Bigg[\sum_{p\,=\,\text{SM\, particles}}
\langle\sigma {\rm v} \rangle_{\eta \eta \rightarrow p\bar{p}}\left( Y_{\eta}^2-(Y_{\eta}^{\rm eq})^2\right) \Bigg]\nonumber \\
&& - \frac{M_{\rm Pl}}{1.66} \frac{x\sqrt{g_{\star}(x)}}{M_{\rm sc}^2\ g_s(x)} \Gamma_{\eta \rightarrow \bar{\psi} \psi}  \ Y_{\eta},
\label{BEeta}
\end{eqnarray}
\begin{eqnarray}
\frac{dY_\psi}{dx} &=& \frac{2 M_{\rm Pl}}{1.66} \frac{x\sqrt{g_{\star}(x)}}{M_{\rm sc}^2 \ g_s(x)} \Gamma_{\eta \rightarrow \bar{\psi} \psi}
\ Y_{\eta}
\label{BEpsi}
\end{eqnarray}
where $x=\dfrac{M_{\rm sc}}{T}$, is a dimensionless variable while
$M_{\rm sc}$ is some arbitrary mass scale which we choose equal to
the mass of $\eta$ and $M_{\rm Pl}$ is the Planck mass. Moreover, $g_s(x)$
is the number of effective degrees of freedom associated to the
entropy density of the Universe and the quantity $g_{\star}(x)$
is defined as
\begin{eqnarray}
\sqrt{g_{\star}(x)} = \dfrac{g_{\rm s}(x)}
{\sqrt{g_{\rho}(x)}}\,\left(1 -\dfrac{1}{3}
\dfrac{{\rm d}\,{\rm ln}\,g_{\rm s}(x)}{{\rm d} \,{\rm ln} x}\right)\,. 
\end{eqnarray}
Here, $g_{\rho}(x)$ denotes the effective number of degrees
of freedom related to the energy density of the Universe at
$x=\dfrac{M_{\rm sc}}{T}$. 
The first term on the right hand side of the Boltzmann equation \eqref{BEeta} corresponds to the self annihilation of $\eta$ into standard model particles and vice versa which play the role in its freeze-out. The second term on the right hand side of this equation corresponds to the dilution of $\eta$ due to its decay into dark matter $\psi$. Let us denote the freeze-out temperature of $\eta$ as $T_F$ and its decay temperature as $T_D$. If we assume that the mother particle freezes out first followed by its decay into dark matter particles, we can consider $T_F > T_D$. In such a case, we can first solve the Boltzmann equation for $\eta$ considering only the self-annihilation part to calculate its freeze-out abundance. \begin{eqnarray}
\frac{dY_{\eta}}{dx} &=& -\frac{4 \pi^2}{45}\frac{M_{\rm Pl} M_{\rm sc}}{1.66}\frac{\sqrt{g_{\star}(x)}}{x^2}\Bigg[\sum_{p\,=\,\text{SM\, particles}}
\langle\sigma {\rm v} \rangle_{\eta \eta \rightarrow p\bar{p}}\left( Y_{\eta}^2-(Y_{\eta}^{\rm eq})^2\right) \Bigg]
\label{BEeta1}
\end{eqnarray}
Then we solve the following two equations for temperature $T < T_F$
\begin{eqnarray}
\frac{dY_{\eta}}{dx} &=&  - \frac{M_{\rm Pl}}{1.66} \frac{x\sqrt{g_{\star}(x)}}{M_{\rm sc}^2\ g_s(x)} \Gamma_{\eta \rightarrow \bar{\psi} \psi}  \ Y_{\eta},
\label{BEeta2}
\end{eqnarray}
\begin{eqnarray}
\frac{dY_\psi}{dx} &=& \frac{2 M_{\rm Pl}}{1.66} \frac{x\sqrt{g_{\star}(x)}}{M_{\rm sc}^2 \ g_s(x)} \Gamma_{\eta \rightarrow \bar{\psi} \psi}
\ Y_{\eta}.
\label{BEpsi2}
\end{eqnarray}
We stick to this simplified assumption $T_F > T_D$ in this work and postpone a more general analysis without any assumption to an upcoming work. The assumption $T_F > T_D$ allows us to solve the Boltzmann equation \eqref{BEeta1} for $\eta$ first, calculate its freeze-out abundance and then solve the corresponding equations \eqref{BEeta2}, \eqref{BEpsi2} for $\eta, \psi$ using the freeze-out abundance of $\eta$ as initial condition\footnote{Recently another scenario was proposed where the dark matter freezes out first with underproduced freeze-out abundance followed by the decay of a long lived particle into dark matter, filling the deficit \cite{Borah:2017dfn}.}. In such a scenario, we can solve the Boltzmann equations \eqref{BEeta2}, \eqref{BEpsi2} for different benchmark choices of $Y, m_{\eta}, m_{\rm DM} \equiv M_{\psi}$ and estimate the freeze-out abundance of $\eta$ that can generate $\Omega h^2 =0.12$, the {\it canonical} value of the dark matter $(\psi)$ relic abundance in the present Universe. This required freeze-out abundance of $\eta$ then restricts the parameters involved in its coupling to the SM particles. It turns out that a scalar singlet like $\eta$ interacts with the SM particles only through the Higgs portal and hence depends upon the $\eta-H$ coupling, denoted by $\lambda_{H \eta}$. In figure \ref{fig1}, we show different benchmark scenarios that give rise to the correct relic abundance of dark matter. In the left panel of figure \ref{fig1}, we show the abundance of both $\eta$ (after its thermal freeze-out) and $\psi$ for benchmark values of their masses $(m_{\eta} = 200 \; \text{GeV}, m_{\rm DM} = 1 \; \text{GeV})$ as a function of temperature for three different values of Yukawa coupling $Y = 10^{-13}, 10^{-12}, 10^{-11}$. It can be clearly seen that while the freeze-out abundance of $\eta$ drops due to its decay into $\psi$, the abundance of the latter grows. The value of $\eta-H$ coupling is chosen to be $\lambda_{H \eta}=0.004$ in order to generate the correct freeze-out abundance of $\eta$ which can later give rise to the required dark matter abundance through its decay. It can be seen that, once we fix the $\psi$ and $\eta$ masses, the final abundance of $\psi$ does not depend upon the specific Yukawa coupling $Y$ as $\eta$ dominantly decays into $\psi$ only. However, different values of $Y$ can lead to different temperatures at which the freeze-in of $\psi$ occurs, as seen from the left panel of figure \ref{fig1}. The right panel of figure \ref{fig1} shows the relic abundance of dark matter for a fixed value of Yukawa coupling $Y = 10^{-12}$ but three different benchmark choices of $m_{\eta}, m_{\rm DM}$ where the $\lambda_{H \eta}$ parameter is chosen appropriately in each case so as to generate the correct freeze-out abundance of $\eta$.
\begin{figure}[h!]
\centering
\includegraphics[width=0.75\textwidth]{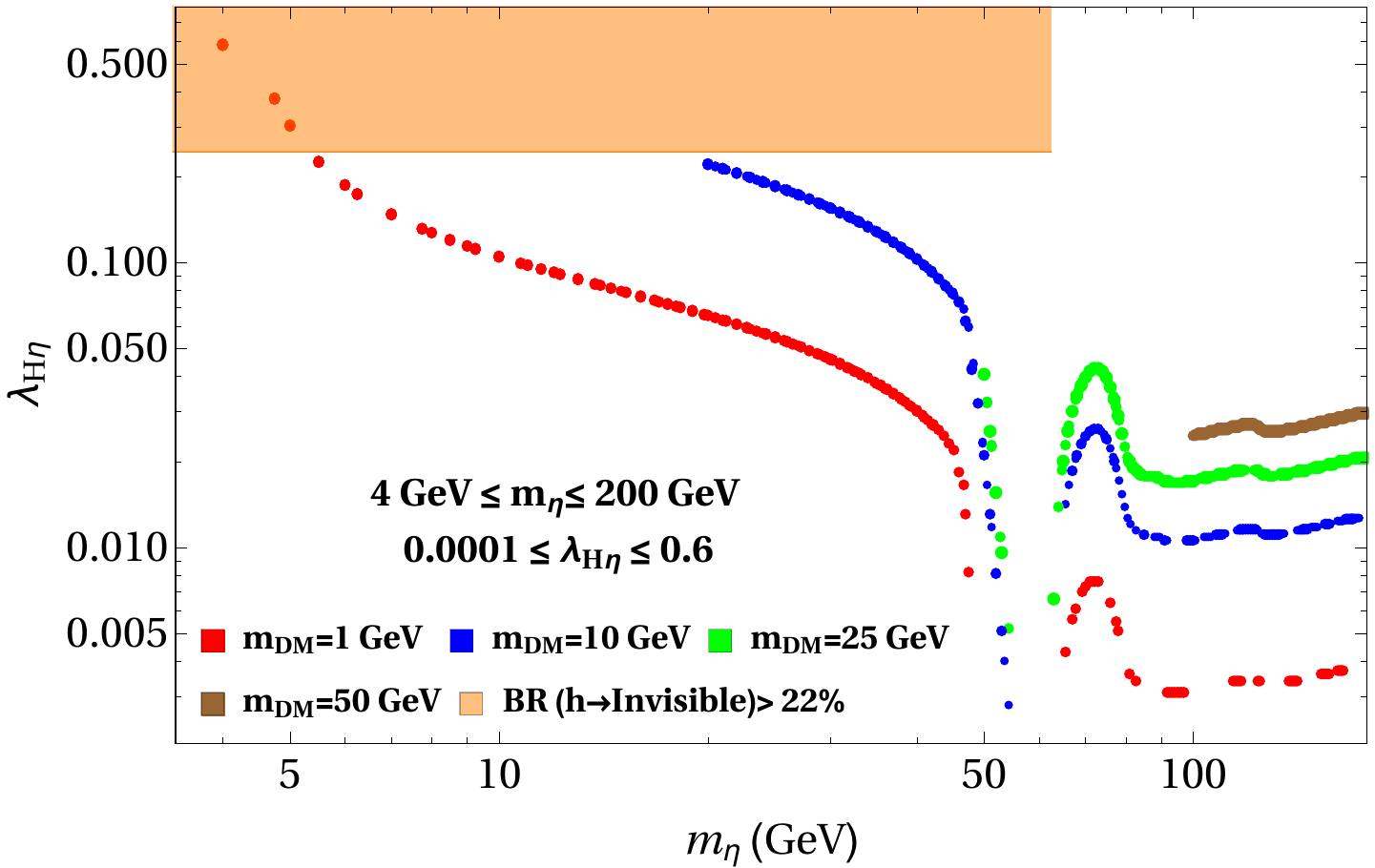}
\caption{The allowed parameter space in $\lambda_{H\eta}$-m$_{\eta}$ plane for different DM masses which gives rise to the required freeze-out abundance of $\eta$ followed by the correct DM abundance from $\eta$ decay.}
\label{fig2}
\end{figure}

\begin{figure}[h!!!!!!]
\includegraphics[width=0.6\textwidth]{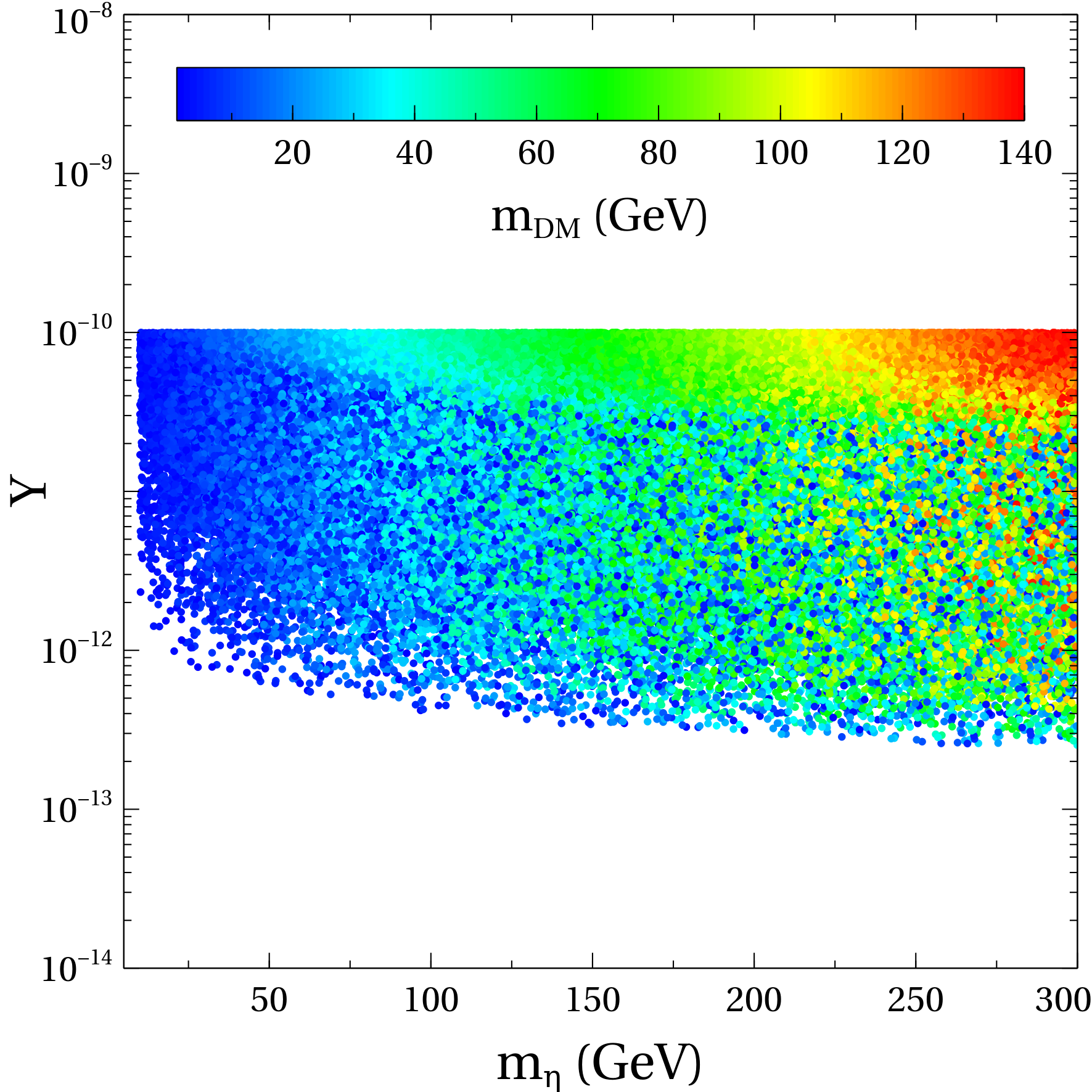}
\caption{Parameter space for Yukawa coupling and $\eta$-DM masses that satisfy the upper and lower bounds on lifetime of $\eta: T_{\rm BBN} < T_D < T_F$.}
\label{fig3}
\end{figure}

The freeze-out abundance of $\eta$ can be calculated similar to the way the relic abundance of scalar singlet dark matter is calculated. For the details of scalar singlet dark matter, one may refer to the recent article \cite{Athron:2017kgt} and references therein for earlier works. In figure \ref{fig2}, we show the parameter space of scalar singlet $\eta$ in terms of $\lambda_{H \eta}, m_{\eta}$ that can give rise to the required freeze-out abundance in order to generate the correct FIMP abundance through $\eta \rightarrow \psi \psi$ decay. In this plot, the resonance region is clearly visible at $m_{\eta} = m_h/2$ where $m_h \approx 125$ GeV is the SM like Higgs boson. The parameter space corresponding to DM mass of 50 GeV is seen only at the extreme right end of the plot in figure \ref{fig2} due to the requirement of $m_{\eta} \geq 2m_{\rm DM}$ to enable the decay of $\psi$ into two DM candidates. Since $\eta$ is long lived and it decays only into DM at leading order, any production of $\eta$ at experiments like the LHC could be probed through invisible decay of SM like Higgs. However, this constraint is applicable only for dark matter mass $m_{\eta} < m_h/2$. The invisible decay width is given by
\begin{equation}
\Gamma (h \rightarrow \text{Invisible})= {\lambda^2_{H \eta} v^2\over 64 \pi m_h} 
\sqrt{1-4\,m^2_{\eta}/m^2_h}
\end{equation}
The latest constraint on invisible Higgs decay from the ATLAS experiment at the LHC is \cite{Aad:2015pla}
$$\text{BR} (h \rightarrow \text{Invisible}) = \frac{\Gamma (h \rightarrow \text{Invisible})}{\Gamma (h \rightarrow \text{Invisible}) + \Gamma (h \rightarrow \text{SM})} < 22 \%.$$
We incorporate this in figure \ref{fig2} and find that some part of parameter space in $\lambda_{H\eta}$-m$_{\eta}$ plane can be excluded for low dark matter masses $m_{\rm DM} < 10$ GeV by LHC constraints.

Although we are considering a simplified case where the decay of mother particle occurs after the mother particle freezes out $T_F>T_D$, we note that this decay can not be delayed indefinitely. Considering the successful predictions of big bang nucleosynthesis (BBN) which occurs around typical time scale $t \sim 1\; \text{s}$, we constrain the lifetime of $\eta$ to be less than this BBN epoch so as not to alter the cosmology post-BBN era. The upper and lower bound on $\eta$ lifetime therefore, constrains the corresponding Yukawa which we show as a scan plot in figure \ref{fig3} for different values of $\eta$ and dark matter masses. Dark matter masses are also varied in such a way that $m_{\eta} \geq 2 m_{\rm DM}$ is satisfied. We have not incorporated the constraints on dark matter relic abundance in figure \ref{fig3}, as we still have freedom in choosing $\lambda_{H \eta}$ that can decide the freeze-out abundance of $\eta$ required for producing correct dark matter abundance through freeze-in. We leave a more general scan of such parameter space to an upcoming work.

It should be noted that we did not consider the production of dark matter from the decay of the flavon $\zeta$ responsible for its mass, as shown in equation \eqref{fimpmass}. Since we intended to explain FIMP coupling and Dirac neutrino mass through same dimension six couplings, we did not take this dimension five term into account. This can be justified if we consider the masses of such flavons to be larger than the reheat temperature of the Universe, so that any contribution to FIMP production from $\zeta$ decay is Boltzmann suppressed. For example, the authors of \cite{Mambrini:2013iaa} considered such heavy mediators having mass greater than the reheat temperature, in a different dark matter scenario. We also note that there was no contribution to FIMP production through annihilations in our scenario through processes like ${\rm SM, SM} \rightarrow \psi \psi$ with $\eta$ being the mediator. This is justified due to the specific flavour transformations of $\eta$ and the fact that $\eta$ does not acquire any vev.

\section{Conclusion}
\label{sec:conc}
We have proposed a scenario that can simultaneously explain the tiny Yukawa coupling required for Dirac neutrino masses from the standard model Higgs field and the coupling of non-thermal dark matter populating the Universe through freeze-in mechanism. The proposed scenario is based on dynamical origin of such tiny couplings from a flavour symmetric scenario based on discrete non-abelian group $A_4$ that allows such couplings at dimension six level only thereby explaining their smallness naturally. The $A_4$ flavour symmetry is augmented by additional discrete symmetries like $Z_N$ and a global lepton number symmetry to forbid the unwanted terms from the Lagrangian. The charged lepton and dark matter masses are generated at dimension five level while the sub-eV Dirac neutrino masses arise only at dimension six level. The correct leptonic mixing can be produced depending on the alignment of flavon vev's. One such alignment which we analyse numerically predicts a normal hierarchical pattern of light neutrino masses and interesting correlations between neutrino oscillation parameters. The atmospheric mixing angle is preferred to be in the lower octant for maximal Dirac CP phase in this scenario.

In the dark matter sector, the effective coupling of non-thermal dark matter ($\psi$, a singlet fermion) with its mother particle ($\eta$, a singlet scalar) arises at dimension six level through the same flavons responsible for neutrino mass. The mother particle, though restricted to decay only to the dark matter particles at cosmological scales, can have sizeable interactions with the standard model sector through Higgs portal couplings. Adopting a simplified scenario where the mother particle freezes out first and then decays into the dark matter particles, we first calculate the freeze-out abundance of $\eta$ and then calculate the dark matter abundance from $\eta$ decay. Although such non-thermal or freeze-in massive particle dark matter remains difficult to be probed due to tiny couplings, its mother particle can be produced at ongoing experiments like the LHC. We in fact show that some part of mother particle's parameter space can be constrained from the LHC limits on invisible decay rate of the SM like Higgs boson, and hence can be probed in near future data. Since $\eta$ is long lived, its decay into dark matter particles on cosmological scales can be constrained if we demand such a decay to occur before the BBN epoch. We find the lower bound on Yukawa coupling $Y$ governing the decay of $\eta$ into DM, and show it to be larger than around $10^{-13}$. We leave a more detailed analysis of such scenario without any assumption of $\eta$ freeze-out preceding the freeze-in of $\psi$ to an upcoming work.

\acknowledgments
DB acknowledges the support from IIT Guwahati
start-up grant (reference number: xPHYSUGIITG01152xxDB001)
and Associateship Programme of IUCAA, Pune. 

\appendix
\section{$A_4$ Multiplication Rules}
\label{appen1}
$A_4$, the symmetry group of a tetrahedron, is a discrete non-abelian group of 
even permutations of four objects. It has four irreducible representations: 
three one-dimensional and one three dimensional which are denoted by $\bf{1}, 
\bf{1'}, \bf{1''}$ and $\bf{3}$ respectively, being consistent with the sum of 
square of the dimensions $\sum_i n_i^2=12$. We denote a generic permutation 
$(1,2,3,4) \rightarrow (n_1, n_2, n_3, n_4)$ simply by $(n_1 n_2 n_3 n_4)$. The 
group $A_4$ can be generated by two basic permutations $S$ and $T$ given by $S = 
(4321), T=(2314)$. This satisfies 
$$ S^2=T^3 =(ST)^3=1$$
which is called a presentation of the group. Their product rules of the 
irreducible representations are given as
$$ \bf{1} \otimes \bf{1} = \bf{1}$$
$$ \bf{1'}\otimes \bf{1'} = \bf{1''}$$
$$ \bf{1'} \otimes \bf{1''} = \bf{1} $$
$$ \bf{1''} \otimes \bf{1''} = \bf{1'}$$
$$ \bf{3} \otimes \bf{3} = \bf{1} \otimes \bf{1'} \otimes \bf{1''} \otimes 
\bf{3}_a \otimes \bf{3}_s $$
where $a$ and $s$ in the subscript corresponds to anti-symmetric and symmetric 
parts respectively. Denoting two triplets as $(a_1, b_1, c_1)$ and $(a_2, b_2, 
c_2)$ respectively, their direct product can be decomposed into the direct sum 
mentioned above. In the $S$ diagonal basis, the products are given as
$$ \bf{1} \backsim a_1a_2+b_1b_2+c_1c_2$$
$$ \bf{1'} \backsim a_1 a_2 + \omega^2 b_1 b_2 + \omega c_1 c_2$$
$$ \bf{1''} \backsim a_1 a_2 + \omega b_1 b_2 + \omega^2 c_1 c_2$$
$$\bf{3}_s \backsim (b_1c_2+c_1b_2, c_1a_2+a_1c_2, a_1b_2+b_1a_2)$$
$$ \bf{3}_a \backsim (b_1c_2-c_1b_2, c_1a_2-a_1c_2, a_1b_2-b_1a_2)$$
In the $T$ diagonal basis on the other hand, they can be written as
$$ \bf{1} \backsim a_1a_2+b_1c_2+c_1b_2$$
$$ \bf{1'} \backsim c_1c_2+a_1b_2+b_1a_2$$
$$ \bf{1''} \backsim b_1b_2+c_1a_2+a_1c_2$$
$$\bf{3}_s \backsim \frac{1}{3}(2a_1a_2-b_1c_2-c_1b_2, 2c_1c_2-a_1b_2-b_1a_2, 
2b_1b_2-a_1c_2-c_1a_2)$$
$$ \bf{3}_a \backsim \frac{1}{2}(b_1c_2-c_1b_2, a_1b_2-b_1a_2, c_1a_2-a_1c_2)$$

\bibliographystyle{apsrev}
\bibliography{ref_fimp.bib}

\end{document}